\font\klein=cmbx10 scaled \magstep1

\font\mittel=cmbx10 scaled \magstep2

\magnification=\magstep1
\baselineskip=16pt
\overfullrule=0pt
\def\vep{\varepsilon}
\def\p{{\bf p}}

\def\x{{\bf x}}
\def\Ln{l_n{\textstyle{\left( {Br^2\over2}\right)}}}

\def\R{[ 1+(M^jr)^N]^{-1}}
\def\T{[ 1+(M^j|\tau|)^{N'}]^{-1}}

\def\sh{{\mathop{\rm{sh}}}}
\def\ch{{\mathop{\rm{ch}}}}
\def\th{{\mathop{\rm{th}}}}

\def\D{\bigtriangleup}

\def\bp{{\bf \Psi}}
\def\bC{{\bf C}}
\def\bS{{\bf S}}
\def\1{{\bf 1}}

\def\G{{\cal G}}
\def\V{{\cal V}}

\def\k{{\bf k}}
\def\bk{{\bf k}}

\def\M{\max_{{c_1\over B}\le n\le {c_2\over B}}
       \Bigl\{ \Bigl|\Ln\Bigr|\Bigr\}}
\def\N{|\!|\!|}
\def\Bbb#1{{\bf #1}}
\def\blacksquare{\vcenter{\vbox{\hrule height .5pt
  \hbox{\vrule width .5pt height 6pt \kern 6pt
        \vrule width .5pt} \hrule height .5pt}}}
\def\gtrapprox{{\textstyle 
      {{\phantom{.}\atop>}\atop\phantom{.}}}\!\llap{${}_{\sim}$}}

\noindent{\mittel I. Introduction}
\bigskip
\bigskip
 In this paper, we consider the model of a many
  electron system in a constant 
    magnetic 
 field in three space time dimensions described by the effective potential
$$\G(\psi^e,\bar\psi^e)=\log{1\over Z}\int 
    e^{-\lambda\V(\psi+\psi^e,\bar\psi+\bar\psi^e)}d\mu_S \eqno(I.1)$$
where the interaction
$${\cal V}(\psi,\bar\psi)=\sum_{\alpha,\beta\in\{\uparrow,\downarrow\}}
   \int d\xi d\xi' \bar\psi_\alpha(\xi)\psi_\alpha(\xi) V(\xi-\xi')
  \bar\psi_\beta(\xi')\psi_\beta(\xi') \eqno (I.2)$$
 is assumed to be  short range and rotation invariant. 
  Here, $d\mu_S$ is the Grassmann Gaussian measure 
   with covariance $S$, where 
$S$ is the exact propagator for a free many electron system in a constant
  magnetic field, 
$$\eqalignno{ S(\x,\tau;\x'\tau')&=e^{i{B\over2}(yx'-xy')}{B\over2\pi} 
     \sum_{n=0}^{\infty} l_n{\textstyle \left( {B|\x-\x'|^2\over2}\right)} 
    e^{-\vep_n (\tau-\tau')}\left[\theta(-\vep_n)\theta(\tau'-\tau)
  -\theta(\vep_n)\theta(\tau-\tau')\right] \cr
 &=e^{i{B\over2}(yx'-xy')}D(\x-\x',\tau-\tau')  & (I.3) \cr}$$
see Lemma II.2. That is, we will consider the interaction
  as a perturbation,
  but the 
 magnetic field is treated exactly without using linear response theory. 
\par
We will prove (see Theorem II.3 for the notation) that 
   the translation invariant part $D$ of $S$  can  be written 
 in momentum space as 
$$D(\k,\tau)=e^{-\vep_{n_B}\tau}
   {1\over \ch^2{B\tau\over2}} \int_0^{\infty} 
   \delta_B(s) e^{- e(\k,s) {\th{B\tau\over 2}\over{B\over2}}}
   \left[\theta(-e(\k,s))\theta(-\tau)
  -\theta(e(\k,s))\theta(\tau)\right] ds $$
$$-e^{-\vep_{n_B}\tau} {2\over 1+e^{-B\tau}}(-1)^{n_B}l_{n_B}{\textstyle 
   \left( {2\k^2\over B}\right) }  \eqno(I.4)$$ 
which may be compared to the free electron
   propagator without magnetic field, 
$$C(\k,\tau)=e^{-e(\k)\tau} \left[\theta(-e(\k))\theta(-\tau)
  -\theta(e(\k))\theta(\tau)\right] \eqno (I.5)$$ 
  Using formula $(I.4)$ and the results of [FT1,2], we derive a 
 BCS-equation with magnetic field (III.2.6) from which the existence of a 
 critical magnetic field follows, see also the curves in Section III. 
\par
As a first step towards rigourosly justifying this BCS-equation and its 
 predictions, we consider  the perturbation theory
 of the model defined by $(I.1)$. We show that graphs containing no two or
 four 
 legged subgraphs are bounded by $const^n$, four
 legged subgraphs produce $n!$'s, $n$ being the order 
  of perturbation theory, 
  and two legged subgraphs have to be renormalized. 
 Furthermore, there is convergence graph by graph to 
  the $B=0$ model. The main  
 problem 
 in proving this is to get the correct propagator 
    estimates. Once this is 
 done, one can apply the machinery of [FT1] to get the stated results. 
\par
 In Section II  
  we compute the magnetic field propagator
   in the symmetric gauge and prove
 formula $(I.4)$.  Since we are interested
   in small magnetic fields, effects  
 comming 
  from the filling factor of the highest occupated
   Landau level are neglected. 
   The BCS-equation with magnetic field is derived 
 in Section III and  Section IV contains a short
   discussion of perturbation  
 theory. 
\par
 In this paper, only the main computations are given.
   The reader who wants
   to see 
   more is refered to [Le], where all calculations 
  are done in great detail. 
\par
 I thank J. Feldman and E. Trubowitz who suggested 
  this interesting problem
   to me 
  and the ETH Z\"urich for the financial support
   during the time this work was  
 done. 
\bigskip
\bigskip
\bigskip\bigskip
\bigskip
\bigskip
\noindent{\mittel II. The Magnetic Field Free Propagator}
\bigskip
\bigskip
\bigskip
The one particle Schr\"odinger equation for
   an electron in a constant magnetic 
 field $\vec B=(0,0,B)$ in two dimensions without spin is 
$$\textstyle 
  H\psi=\left\{ {1\over 2m} \left( {\hbar\over i}\nabla-eA\right)^2-\mu\right\}\psi=
   \vep\psi \eqno (II.1) $$
where $\mu$ denotes the chemical potential.  The propagator $S$
   will be calculated in the symmetric gauge 
$$\textstyle A(x,y)=\left(-{B\over2}y,
   {B\over2}x,0\right) \>.\eqno (II.2)$$
A computation in a nonsymmetric gauge 
  $\tilde A(x,y)=\left( -By,0,0\right)\>$
  is given in [Le]. 
 In the first case, the eigenfunctions of $(II.1)$ are 
  labelled by two discrete
   parameters 
 whereas in the second case one gets one discrete and one
   continous parameter.
  So in the
  first case the calculation of the covariance involves
   two infinite sums 
  and in the
 second case there is one integral and one infinite sum
  which of course yield the same result up to the phase factor 
$$S(\xi,\xi')= e^{i\left({B\over2}xy
  -{B\over2}x'y'\right)}\tilde S(\xi,\xi')\>
    \eqno (II.3)$$
  which is due to the gauge transformation 
$$\textstyle A(x,y)=\tilde A(x,y)+\nabla
  \left( {B\over2}xy\right)\>. \eqno (II.4)$$
Here $\xi=(\x,\tau)=(x,y,\tau)$. 
\par
For simplicity, the computation is done in infinite
   volume and it is assumed  
  that the 
  highest occupated Landau level is fully occupated
   with electrons, that is
   (see  $(II.6)$
  below) 
 $\vep_n\ne 0\;\forall n\in \Bbb N\>,$  so effects comming 
 from the filling factor of the highest Landau level
   are neglected. This should
 be no 
 restriction in considering magnetic fields
   concerning superconductivity,
  because there
 the number of Landau levels is of order $10^4$ or $10^5$ (for BCS  
  superconductors). 
\par
We now present the calculation of $S$ in the symmetric gauge.
  In Section III and IV  all calculations are done with $S$. 
  The following lemma summarizes the properties 
  of the eigenfunctions of $(II.1)$ which are well known. 
\bigskip
\bigskip
\noindent{\bf Lemma II.1 (Eigenfunctions):} Put $e,\hbar$ and the 
 electron mass  to one 
 and identify $(x,y)$ with $z=x+iy$. Then the 
 normalized eigenfunctions of $(II.1)$ in the symmetric gauge 
 $(II.2)$ are given  by 
$$\textstyle 
  \phi_{nm}(z)=\left({B\over2\pi}\right)^{1\over2} \left({n!\over  
  m!}\right)^{1\over2}
   \left( \sqrt{B\over2} z\right)^{m-n} L_n^{m-n}
   \left( {B|z|^2\over 2}\right) 
   e^{-{B|z|^2\over4}} \eqno (II.5)$$
with energy eigenvalues
$$\textstyle \vep_{nm}=\vep_n= {B}\left( n+{1\over2}\right)
  -\mu \eqno (II.6)$$
 where $n,m\in\{0,1,2,\cdots\}$ and $m-n$ has the
   meaning of angular momentum. 
 They have the following properties: \par
$$\eqalignno{ \phi_{nm}(z)^* &=(-1)^{n-m}\phi_{mn}(z)  & (II.7a)  \cr
  \hat \phi_{nm}(k=k_1+ik_2)\phantom{^*}&
  =\textstyle{{4\pi\over B}}(-1)^n \phi_{nm}\textstyle{\left( -{2\over 
   B}ik\right)}
    & (II.7b)  \cr
 \sum_{m=0}^{\infty} \phi_{n_1 m}(z_1)\phi_{n_2 m}(z_2)^* &=\textstyle{
    \left( {B\over2\pi} 
   \right)^{1\over2}} \phi_{n_1 n_2}(z_1-z_2)
    e^{i{B\over2}Im(z_1z_2^*)} \>.
   & (II.7c)\cr} $$
\goodbreak
\bigskip
\bigskip
\noindent {\bf Lemma II.2 (Propagator):} The magnetic field 
  free propagator with imaginary time in infinite volume 
  and symmetric gauge is 
given by
$$ S(\xi,\xi')=\sum_{n,m=0}^{\infty} \phi_{nm}(z)\phi_{nm}(z')^*
   e^{-\vep_n (\tau-\tau')}[\vep_n,\tau-\tau']  $$
$$= e^{i{B\over2}(yx'-xy')}D(\xi-\xi')  \eqno (II.8a)$$
where the translation invariant part $D$ of $S$ is given by
$$D(\xi)={B\over2\pi} \sum_{n=0}^{\infty} l_n{\textstyle \left( 
   {Br^2\over2}\right)} 
    e^{-\vep_n \tau}[\vep_n,\tau]\>.\eqno (II.8b)$$
Here, $l_n(v)=L_n(v)e^{-{v\over2}}$ denotes the Laguerre function, 
$$[\vep_n,\tau]=\left[\theta(-\vep_n)\theta(-\tau)
  -\theta(\vep_n)\theta(\tau)\right]=\cases{ -1& if 
      $\vep_n>0\>\wedge \>\tau>0$ \cr
         1& if $\vep_n<0\>\wedge\> \tau<0$ \cr
           0& else  \cr} \eqno(II.8c) $$
and  $\vep_n=
  B\Bigl(n+{1\over2}\Bigr)-\mu$. The spatial
   Fourier transform of $D$ is
 $$D(\bk,\tau)=\sum_{n=0}^{\infty} 2(-1)^n 
   l_n{\textstyle \left( {2\bk^2\over
     B}\right)} 
    e^{-\vep_n\tau}[\vep_n,\tau] \eqno(II.8d)$$\nobreak
where $\bk=(k_1,k_2)$. 
\bigskip
\noindent {\bf Proof:} By definition (see for example [FW]),
   the imaginary time 
 propagator is 
$$S(\xi,\xi')=\sum_{nm}\phi_{nm}(z)\phi_{nm}(z')^*
    e^{-\vep_n (\tau-\tau')} 
 \left[\theta(-\vep_n)\theta(\tau'-\tau)-\theta(\vep_n)
      \theta(\tau-\tau')\right] $$
where $\theta(v)$ denotes the step function
   which is one for $v>0$ and zero 
 otherwise. 
 Use $(II.7c)$ to perform the $m$-sum:
$$\eqalignno{ \sum_{m=0}^{\infty} \phi_{nm}(z)\phi_{nm}(z')^* &=
  {B\over 2\pi} L_n{\textstyle \left( {B|z-z'|^2\over2}\right)}
      e^{-{B|z-z'|^2\over4}} 
    e^{i{B\over2}(yx'-xy')}\>. &  \cr}$$
The Fourier transform is computed with $(II.7b)$ for $n=m$:
$$\left[ l_n{\textstyle \left( {Br^2\over2}\right)}
   \right]^{\wedge{}}(\bk)
     ={2\pi\over B} 2(-1)^n
   l_n{\textstyle \left({2\bk^2\over B}\right)} \, .$$
thus $(II.8c)$ follows $\blacksquare$
\bigskip
\bigskip
The free $B=0$ propagator in the mixed representation,
    that is in $(\bk,\tau)=
 (k_1,k_2,\tau)$-space, 
is given by
$$C(\bk,\tau)=e^{-e(\bk)\tau} [e(\bk),\tau] \eqno (II.9) $$
with
$$e(\bk)={\bk^2\over2}-\mu. \eqno (II.10)$$
It is hard to see from $(II.8d)$, how $(II.9)$ is obtained as $B$ goes 
  to zero. From the following representation $(II.11)$
   of $D$ the limit $B\to 0$ can be read off. 
The main use of formula $(II.11)$ below is that it simplifies 
  the evaluation of Feynman graphs in 
 momentum space, in particular, the computation of the critical
magnetic field in Section III. 
\bigskip
\bigskip
\noindent {\bf Theorem II.3 (Propagator):} In $(\bk,\tau)$-space, 
  the translation invariant part 
   of the 
 magnetic field free propagator is given by 
$$D(\bk,\tau)=e^{-\vep_{n_B}\tau}
   {1\over \ch^2{B\tau\over2}} \int_0^{\infty} 
   \delta_B(s) e^{- e(\bk,s) {\th{B\tau\over 2}\over{B\over2}}}
   [e(\bk,s),\tau] ds $$
$$-e^{-\vep_{n_B}\tau} {2\over 1+e^{-B\tau}}(-1)^{n_B}l_{n_B}{\textstyle 
   \left( {2\bk^2\over B}\right) }  \eqno(II.11)$$
where ${\rm ch},\>{\rm th}$ are the hyperbolic cosine, tangent,  
  $l_n(v)=L_n(v)e^{-{v\over2}}$
  is the Laguerre function, 
   $\;e(\bk,s)={\bk^2\over2}-\mu s,\;\;\vep_{n_B}
  =B\left(n_B+{1\over2}\right) 
  -\mu$
and 
 $$n_B=\left[ 
   {\mu\over B}+{1\over2}\right] \eqno (II.12)$$
 where the square brackets in $(II.12)$ are Gauss brackets, 
   thus $0\le \vep_{n_B} \le B$.  Furthermore, 
$$\delta_B(s)=2{\mu\over B}(-1)^{n_B}l_{n_B}\left( 4{\mu\over B} s\right)
  \eqno (II.13)$$
  is a 
  $\delta$-sequence with limit $\delta(s-1)$.
\bigskip
\noindent {\bf Remark:}  The second term on the right
   hand side of $(II.11)$   
  converges 
  pointwise to zero, since for 
 $s\ge\epsilon>0$ there is the estimate
      $|l_n(s)|\le{const\over n^{1\over 4}}$ 
 and $n_B$ goes to infinity if $B$ goes to zero. 
\bigskip  
\noindent {\bf Proof:}  $n_B$ is by definition the 
  smallest natural number such 
 that $\vep_{n_B}>0$. Thus
$$\eqalignno{ D(\k,\tau)= &\,2e^{-{B\over2}\tau+\mu\tau}
  \sum_{n=0}^{n_B-1}
    (-1)^n l_n{\textstyle \left( {2\bk^2\over B}\right)} 
    e^{-Bn\tau}\theta(-\tau)\cr
  & -2e^{-{B\over2}\tau+\mu\tau}\sum_{n=n_B}^\infty
  (-1)^n l_n{\textstyle \left( {2\bk^2\over B}\right)}
   e^{-Bn\tau}\theta(\tau)\>. \cr}$$
We will now prove the formulae:
$$\sum_{k=0}^{n-1}  (-1)^k l_k(x) t^k= 
  {t\over (1+t)^2} t^n \int_x^{\infty} ds 
  (-1)^n l_n(s)e^{-{t-1\over t+1}{s-x\over 2}} 
  -{1\over 1+t} t^n (-1)^n l_n(x)
   \eqno (II.14) $$\nobreak
and for $|t|<1$ it is\nobreak
$$\sum_{k=n}^{\infty} (-1)^k l_k(x) t^k
  ={t\over (1+t)^2} t^n \int_0^x ds
    (-1)^n l_n(s) e^{-{1-t\over 1+t} {x-s\over2}} 
  + {1\over 1+t} t^n (-1)^n 
    l_n(x)\>.
   \eqno (II.15)$$
The following proof is short, but one has already
    to know the answer. An 
  alternative proof which makes clear how  the above 
formulae have been found is given in [Le]. 
 \par
For $|t|<1$, let 
$$G(x,t)={1\over 1+t} e^{-{1-t\over1+t}{x\over 2}}
  =\sum_{k=0}^\infty l_k(x) 
     (-t)^k$$
be the generating function for the Laguerre functions. We have 
$$DG(x,t)=0$$
where
$$D=\Bigl\{ 2(1+t){d\over dx}+(1-t)\Bigr\}\,.$$
Hence
$$2l_0'(x)+l_0+\sum_{k=1}^\infty \left( 
     2l_k'(x)-2l_{k-1}'(x)+l_k(x)+l_{k-1}(x)\right) 
   (-t)^k=0$$
which gives the recursion relation 
$$2l_0'+l_0=0,\;\;\;\;2l_k'-2l_{k-1}'+l_k+l_{k-1}=0,\;\;k\ge 1\,.$$
 Now let $s_n(x,t)$ be the left hand side and $i_n(x,t)$ be 
    the right hand side 
 of $(II.14)$. Then
$$\eqalignno{ Ds_n(x,t)&=2l_0'(x)+l_0(x)+\sum_{k=1}^{n-1}
  \left( 2l_k'(x)-2l_{k-1}'(x)+l_k(x)+l_{k-1}(x)\right) 
   (-t)^k \cr
  &\phantom{=}-2l_{n-1}'(x)(-t)^n+l_{n-1}(x)(-t)^n \cr
  &=\left(-2l_{n-1}'(x)+l_{n-1}(x)\right) (-t)^n \cr}$$
and
$$\eqalignno{Di_n(x,t)&=2{-t\over 1+t}(-t)^n l_n(x)-2(-t)^n
     l_n'(x)-{1-t\over 1+t}(-t)^nl_n(x) \cr
  &=-\left(l_n(x)+2l_n'(x)\right) 
   (-t)^n=-\left(2l_{n-1}'(x)-l_{n-1}(x)\right)(-t)^n
   =Ds_n(x,t) \cr}$$
where in the last line the recursion relation has
   been used. Therefore, the difference 
  $\Delta_n(x,t)=s_n(x,t)-i_n(x,t)$ obeys $D\Delta_n(x,t)=0$ which gives 
$$\Delta_n(x,t)=\Delta_n(0,t)\>e^{-{1-t\over1+t}{x\over 2}}\,.$$
But ([GR],7.414.6)
$$\int_0^\infty ds\>(-1)^n l_n(s) e^{-{t-1\over t+1}{s\over2}} =
   {1+t\over t}{1\over t^n}     $$
thus
$$\eqalignno{ i_n(0,t)&={t\over (1+t)^2} t^n\int_0^\infty
   ds\,(-1)^nl_n(s)
     e^{-{t-1\over t+1}{s\over2}}  -{1\over 1+t} t^n(-1)^nl_n(0)  \cr
 &={1\over 1+t}-{1\over 1+t} (-t)^n  \cr
 &=\sum_{k=0}^{n-1}(-t)^k=s_n(0,t) \cr}$$
which proves $(II.14)$. Formula $(II.15)$ is obtained by writing
$$\sum_{k=n}^{\infty} (-1)^k l_k(x) t^k={1\over 1+t} 
  e^{-{1-t\over1+t}{x\over 
   2}}-
   \sum_{k=0}^{n-1}  (-1)^k l_k(x) t^k$$
and
$${t\over (1+t)^2} t^n \int_0^x ds
    (-1)^n l_n(s) e^{-{1-t\over 1+t} {x-s\over2}}=$$
$$ {1\over 1+t} e^{-{1-t\over1+t}{x\over 2}}-
  {t\over (1+t)^2} t^n \int_x^{\infty} ds 
  (-1)^n l_n(s)e^{-{t-1\over t+1}{s-x\over 2}}.$$
Now put $t=e^{-B\tau}$, $x={2\k^2\over B}$ and $n=n_B$ in $(II.14,15)$ 
 and substitute the integration variable $s$ by $v={B\over 4\mu}s$
   to obtain the final 
 result $(II.11)$.  A detailed 
  discussion of the $\delta$-sequence is given in [Le]. See also
   the curves below
   $\blacksquare$
\bigskip
\bigskip
\noindent$\phantom{m}$
In comparing $(II.8,11)$ with $(II.9)$, one can see 
  essentially three differences.
\item{(i)} The magnetic field propagator is no longer
   translation invariant.   
  This is 
 due to the fact, that  linear momentum is no
   longer an eigenstate  but 
  angular momentum. 
\item{(ii)} In momentum space, there is no longer
   a sharp fermi surface (or  
  Fermi
 circle, since we are in two dimensions). Rather,
   the Fermi surface is smeared  
   out with 
 the delta sequence $\delta_B(s)$ so that the density of states 
  in momentum space 
 $\theta\Bigl(\mu-{\bk^2\over 2}\Bigr)$ is substituted by
    $\int_0^{\infty}ds\>
 \delta_B(s)\theta\Bigl(\mu s-{\bk^2\over 2}\Bigr)=
   1-\int_0^{\bk^2\over2\mu}ds\>
 \delta_B(s)$, see the following curves.
\bigskip
$$ (see\;figure\;in\;CMP\;paper)$$
\bigskip
\item{(iii)} The imaginary time variable $\tau$ is substituted by 
  ${th{B\tau\over2}\over {B\over2}}$. This is the most important
    effect (for our purposes), since it 
 changes significantly the values of the Feynman graphs.
   Graphs containing two legged subgraphs become
 finite and the flow for the four point function is expected to
   be convergent  for 
 $B\gtrapprox e^{-{const\over\lambda}}$, see Section III.3.
   Furthermore,
    the  factor
 ${th{B\tau\over2}
  \over {B\over2}}$ appears directly in the BCS-equation with magnetic
  field  $(III.2.6)$,
 and is responsible there for the existence of a critical
   magnetic field. To see this 
  factor, it is necessary to compute the infinite sum over the 
  Laguerre  polynomials, 
 since, mathematically, it comes from the generating function for
   the Laguerre polynomials. Physicially, it expresses the
   fact that electrons are localized by a 
  magnetic field. 
\bigskip
\noindent$\phantom{m}$
Finally, the factor $e^{-\vep_{n_B}\tau}$ requires 
  a short discussion. The 
 assumption that the highest Landau level is fully
   occupated with electrons 
 means that there is no $n\in \Bbb N$ such that $\vep_n=\nobreak 0$. 
  $n_B$ is by  definition the 
 smallest natural number such that $\vep_{n}>0$. Then $0<\vep_{n_B}<B$,
  so one  can write  
$$\vep_{n_B}=\alpha B \eqno (II.16a)$$
 with some $0<\alpha<1$. However, for $\alpha$ arbitrary 
 close to 0 or $B$, the $\tau$-decay of the second
   term in $(II.11)$ becomes 
 arbitrary bad, although it converges pointwise to zero.
   In order to keep the 
 estimates of the following Sections (for example the estimate for
  the second  term 
    in the 
  BCS-equation $(III.2.6)$) uniform, assume 
$$\epsilon\le \alpha\le 
 1-\epsilon \eqno (II.16b)$$
 or equivalently define the set of admitted magnetic fields to be
$${\cal B}=\bigcup_{
  n\in \Bbb N} \left[ {\mu\over n+{1\over2}-\epsilon},
  {\mu\over n-{1\over2}+
   \epsilon} \right] \eqno (II.16c)$$
Then the measure of the set of neglected fields can be made
  arbitrarily small since 
$$\sum_{n=1}^{\infty}\biggl( {\mu\over n+{1\over2}-\epsilon}
    -{\mu\over n+{1\over2}
   +\epsilon} \biggr) =2\mu\epsilon\sum_{n=1}^{\infty} 
  {1\over \left(n+
   {1\over2}\right)^2+\epsilon^2}\le const\>\epsilon.$$
\eject
\noindent{\mittel III. The Existence of the Critical Field}
\bigskip
\bigskip
\noindent{\klein III.1 The $B=0$ BCS-Equation}
\bigskip
 In [FT2], Feldman and Trubowitz obtained the BCS-equation (without 
 magnetic field)  in the following way: \par
 Consider the effective potential for an interacting many electron
  system which is 
 given by
$$\G(\psi^e,\bar\psi^e)=\log{1\over Z}  \int
  e^{-\lambda\V(\psi+\psi^e,\bar\psi+\bar\psi^e)}d\mu_C(\psi,\bar\psi)
    \eqno (III.1.1)$$
where $C$ is the free propagator corresponding to
   the normal ground state
$$C(\xi,\xi')=C(\xi-\xi')=\int{d^2k\over(2\pi)^2} e^{i\k(\x-\x')}
   e^{-e(\k)(\tau-\tau')}\left[\theta(-e(\k))\theta(-\tau)-
   \theta(e(\k))\theta(\tau)\right]$$
$$=\int{d^2k\over(2\pi)^2} \int{dk_0\over 2\pi} 
  e^{i\k(\x-\x')-ik_0(\tau
   -\tau')} {1\over ik_0-e(\k)}   \eqno (III.1.2)$$
and $e(\k)={\k^2\over2}-\mu$. The quartic interaction ${\cal V}$
   is assumed to be short
ranged. They started to analyse $\G$ in perturbation theory [FT1].
   It turned out that all
graphs containing no two or four legged subgraphs 
  are bounded by $const^n$,
   $n$ 
 denoting the order of perturbation theory, that all graphs which 
  contain no two legged
 subgraphs are bounded by $n!\>const^n$, and that graphs containing
   two legged subgraphs
 are in general infinite.  They introduced a localization operator $L$,
   which acts 
 nontrivially only on quadratic and quartic monomials and isolates
   the  singularities 
 and $n!$'s produced by the two and four legged subgraphs. Then 
  all graphs  contributing 
 to $(1-L)\G$ are bounded by $const^n$. In [FMRT] 
  and [FKLT1,2] it is shown  
  that $(1-L)\G$ is indeed an analytic function at $\lambda=0$ with a 
 fixed, volume independent positive radius of convergence.
\par
 The relevant part $L\G$ of 
 the effective potential is analyzed by a 
  renormalization group flow [FT2]. 
  If one expands 
 the kernel $F^{(h)}$ of the quartic part of $L\G^{(h)}$ into  a 
 Fourier series, 
$$F^{(h)}({ t}',{ s}')=F^{(h)}(\cos\theta)=\sum_{\ell=0}^\infty
  \lambda_\ell^{(h)}\cos\ell\theta\>,$$
 where $k'=(0,k_F{\textstyle {\bf k\over\|\bf k\|}})$ denotes
  the projection  onto 
 the Fermi surface,  one obtains in the
 ladder approximation the following flow equation for the coefficients
 $\lambda_\ell^{(h)}$, see [FT2]:
$$\lambda_\ell^{(h-1)}=\lambda_\ell^{(h)}+\beta^{(h)}
    (\lambda_\ell^{(h)})^2,\;\;\;\ell\ge 0
   \eqno (III.1.3)$$
where 
$$\beta^{(h)}=\int{dp_0d|\p|\over (2\pi)^{3}}  {|\p|\over k_F}
   \left( |C^{(\le h)}(p)|^2-|C^{(<h)}(p)|^2\right)  $$
$$=\int {d|\p|\over (2\pi)^2}  {|\p|\over k_F} 
  {1\over 2|e(\p)|} \left[ \rho^2\left( M^{-2(h+1)}e(\p)^2\right)-
   \rho^2\left( M^{-2h}e(\p)^2\right) \right] \eqno (III.1.4a)$$
approaches the limit 
$$\beta={1\over (2\pi)^2 k_F} \int_0^{\infty} dy {1\over y} \left[ \rho^2(M^{-2}y^2)-
   \rho^2(y^2)\right] \eqno (III.1.4b)$$
and $\rho(x)$ is some $C_0^{\infty}$ function, which is one if $x\le 1$ 
  and zero  if 
 $x\ge M^2$, $M$ being some constant bigger than one. If one starts with
 $\lambda_\ell^{(0)}>0$, which corresponds to an attractive potential,
  then the sequence 
 generated by $(III.1.3)$ diverges to infinity which is 
  interpreted as: the normal 
  ground state is not stable. In order to get a well defined
    effective potential, one 
 introduces a $\D$ in the following way:
\par
Define the two component Nambu fields
$$\bp(\xi)=\left( {\psi_{\uparrow}(\xi)\atop \bar\psi_{\downarrow}(\xi)}\right),\;\;\;
   \bar\bp(\xi)=\left(\bar\psi_{\uparrow}(\xi),
  \psi_{\downarrow}(\xi)\right) 
   \eqno (III.1.5a)$$
or in momentum space 
$$\bp(\k,k_0)=\left( {\psi_{\uparrow}(\k,k_0)
     \atop \bar\psi_{\downarrow}(-\k,-k_0)}
   \right),\;\;\; \bar\bp(\k,k_0)=\left(
   \bar\psi_{\uparrow}(\k,k_0),\psi_{\downarrow}(-\k,-k_0)\right)\>. 
   \eqno (III.1.5b)$$
Then the Grassman Gaussian measure becomes $d\mu_{\bC}(\bp,\bar\bp)$
  with covariance
 matrix 
$$\bC(\xi,\xi')=\big< \bp(\xi)\bar\bp(\xi')\big> 
  =\int {d^{3}k\over (2\pi)^{3}}
   e^{i\k(\x-\x')-ik_0(\tau-\tau')} (ik_0\1 -e(\k) \sigma^3)^{-1}
   \eqno (III.1.6)  $$
 and  the effective potential can be written formally as 
$$\G(\bp^e,\bar\bp^e)=\log{1\over Z} \int
   e^{-\lambda\V(\bp+\bp^e,\bar\bp+\bar\bp^e)}e^{- 
   \int {d^{3}k\over(2\pi)^{3}} \bar\bp(k)
  \left( ik_0\1-e(\k) \sigma^3\right) 
   \bp(k) } d(\bp,\bar\bp)$$
Then add and subtract $\int {d^{3}k\over(2\pi)^{3}}
    \bar\bp(k) \D\sigma^1\bp(k)$
  to get 
$$\eqalignno{\G(\bp^e,\bar\bp^e)&=\log{1\over Z} \int
   e^{-\lambda\V(\bp+\bp^e,\bar\bp+\bar\bp^e)-\D
  \int {d^{3}k\over(2\pi)^{3}} \bar\bp(k) \sigma^1\bp(k) }\times  \cr
 & \cr
 &\phantom{==}e^{-\int {d^{3}k\over(2\pi)^{3}} \bar\bp(k)
      \left( ik_0\1-e(\k) \sigma^3
   -\D\sigma^1\right) 
   \bp(k) } d(\bp,\bar\bp)  \cr
 & \cr
 &=\log{1\over Z'} \int e^{-\lambda\V(\bp+\bp^e,\bar\bp+\bar\bp^e)
    -\D\int d\xi\bar\bp(\xi)\sigma^1\bp(\xi)} 
   d\mu_{\bC_{\D}} & (III.1.7) \cr} $$
where now 
$$\bC_{\D}(\xi,\xi')=\int {d^{3}k\over(2\pi)^{3}} e^{i\k(\x-\x')-ik_0
   (\tau-\tau')} \left( ik_0\1-e(\k)\sigma^3
  -\D\sigma^1\right)^{-1} \eqno (III.1.8)$$
\par
The new covariance is bounded by ${const\over\D}$, which has 
  the consequence that now 
 all graphs are finite.  But if $\D$ is going to zero, graphs 
 containing two  legged 
 subgraphs diverge. To produce an expansion uniform in $\D$ one has to
 renormalize, that is, one has to add counterterms 
$$\delta\V=\delta\mu(\lambda,\mu,\D)\int d\xi\>\bar\bp(\xi)
    \sigma^3\bp(\xi) \eqno (III.1.9)$$
$${\cal D}=D(\lambda,\mu,\D)\int d\xi\>\bar\bp(\xi)\sigma^1\bp(\xi)
      \eqno (III.1.10)$$
 to the exponent  in $(III.1.7)$. But then, to recover the physicial 
 effective potential, one has to impose the constraint
$$\D=-D(\lambda,\mu,\D) \eqno (III.1.11)$$
which, in first order, gives the BCS equation:
\vskip 0.3cm
$$\D=-D(\lambda,\mu,\D)=
     {\lambda\over2} Tr[\sigma^1(\;first\;order\;graphs\;)|_{
  k_0=0,|\k|=k_F} ]$$
$$\eqalignno{ &=-{\lambda\over2} Tr[\sigma^1
    \int{d^{3}p\over (2\pi)^{3}} \big<k,p|
   V|p,k\big>\sigma^3\bC_{\D}(p)\sigma^3  |_{k_0=0,|\k|=k_F} ] \cr
 &=-\lambda \int{d^{3}p\over (2\pi)^{3}}
   \big<k',p| V|p,k'\big> {\D\over p_0^2+
   {e(\p)}^2+\D^2}| & (III.1.12) \cr}$$
where $k'=\left( 0,k_F{\k\over |\k|}\right)$.
   Taking $\big<k',p| V|p,k'\big>
  =\theta(\omega_D-|e(\k)| )\theta(\omega_D-|e(\p)| )$, one obtains the 
 familiar equation 
$$\eqalignno{ 1&=-\lambda\>const \int d^2p \>\theta(\omega_D-|e(\p)| )
   {1\over\sqrt{{e(\p)}^2
   +\D^2}}  \cr
  &=-\lambda\>const \int_0^{\omega_D\over \D} dv {1\over \sqrt{v^2+1}}\>.
    &(III.1.13) \cr}$$
 which gives $\triangle=\omega_D\> e^{-{const\over\lambda}}\>.$
\bigskip
\bigskip\goodbreak
\noindent{\klein III.2 The BCS-Equation with Magnetic Field}
\bigskip
In the case with magnetic field, one can proceed in an analogous way.
  However, because 
 the Hamiltonian is diagonal in $(n,m)$-space, one has to work 
 in this space rather 
than in momentum space. In $(n,m,k_0)$-space, the covariance is 
$$S(n,m,k_0)={1\over ik_0-\vep_n}\>. \eqno (III.2.1)$$
\par
Introduce the two component fields $(III.1.5)$.
  Since $\phi_{nm}(z)^*=(-1)^{n-m}\phi_{ mn}(z)$, 
   the two component fields in $(n,m)$-space  are given by 
$$\eqalignno{ \bp(n,m,k_0)&=\left({ \psi_{\uparrow}(n,m,k_0)\atop  
  (-1)^{n-m} \bar\psi_{\downarrow}
   (m,n,-k_0)}\right), \cr
 & \cr
 \bar\bp(n,m,k_0)&=\left (\bar\psi_{\uparrow}(n,m,k_0),(-1)^{n-m} 
   \psi_{\downarrow}(m,n,-k_0) \right)  & (III.2.2) \cr}$$
and the covariance matrix becomes 
$$\bS(n,m,k_0)=\left( ik_0\1
  -\vep_n\sigma^3\right)^{-1}\>. \eqno (III.2.3)$$
Then writing the integration measure formally as 
$$\exp\left\{ -\sum_{n,m=0}^{\infty} \int{dk_0\over2\pi}
\bar\bp(n,m,k_0)(ik_0\1-\vep_n 
   \sigma^3)\bp(n,m,k_0) \right\} d(\bp,\bar\bp)$$
and adding and subtracting the term 
$$\sum_{n,m=0}^{\infty} 
  \int{dk_0\over2\pi} \bar\bp(n,m,k_0)\D\sigma^1\bp(n,m,k_0)
 \eqno (III.2.4)  $$
one obtains the new covariance 
$$\bS_{\D}(n,m,k_0)=\bS_{\D}(n,k_0)=
    \left( ik_0\1-\vep_n\sigma^3-\D\sigma^1\right)^{-1} $$
which becomes in coordinate and momentum space
$$\eqalignno{ \bS_{\D}(\xi,\xi')&=
   \sum_{n,m=0}^{\infty} \phi_{nm}(z)\phi_{nm}(z')
   \int{dk_0\over2\pi} e^{-ik_0(\tau-\tau')} 
  \left( ik_0\1-\vep_n\sigma^3-\D\sigma^1\right)^{-1} \cr
 & \cr
 &=e^{i{B\over2}\x{{\x}'}^{\bot}} {\bf D}_{\D}(\xi-\xi') &  \cr} $$
where $\x{{\x}'}^{\bot}=yx'-xy'\>, $
$${\bf D}_{\D}(\xi)={B\over2\pi} \sum_{n=0}^{\infty} l_n{\textstyle 
    \left( {Br^2\over2}\right) }  
   \int{dk_0\over2\pi} e^{-ik_0(\tau-\tau')} 
  \left( ik_0\1-\vep_n\sigma^3-\D\sigma^1\right)^{-1}\>, $$
$${\bf D}_{\D}(\k,k_0)=\sum_{n=0}^{\infty} 2(-1)^n l_n{\textstyle
   \left( {2\k^2\over
   B}\right)} \left( ik_0\1-\vep_n\sigma^3-\D\sigma^1\right)^{-1}\>.  $$
Then the BCS-equation with magnetic field is given by the first
  order approximation 
 to the constraint 
$$\D=-D(\lambda,\mu,B,\D)\>. \eqno (III.2.5)$$
\bigskip
\noindent{\bf Theorem III.2.1 (BCS-Equation with Magnetic Field):} Let
   $\big<k,p| V|p,k\big>
  =\phantom{mm}$ $\theta(\omega_D-|e(\k)| )\>
  \theta(\omega_D-|e(\p)| )$. Then, using the 
 approximation $\delta_B(s)\approx \delta(s-1)$, the first order 
 approximation to the constraint $(III.2.5)$ is given by the equation 
$$1=const \>\lambda \int_0^{\infty}
   dt\>J_0(t)\biggl\{\ch\alpha{\textstyle{{B\over\D}}}t 
   {{B\over\D}
   \over \sh{B\over\D}t} \left( 1-e^{-2{\omega_D\over B}
   \th{Bt\over 2\D}}\right) 
  +{1\over2}{B\over \D} {\sh\left[\left({1\over2}
    -\alpha\right) {B\over\D}t\right]
     \over \ch {1\over2}{B\over \D}t}\biggr\}
  \eqno (III.2.6)$$
where $J_0$ denotes the zeroth Bessel 
  function and $\alpha$ is defined by 
   $\vep_{n_B}=\alpha B$, $\epsilon\le 
  \alpha \le 1-\epsilon,$  see $(II.16)$. 
\bigskip
\noindent  {\bf  Remarks:} {\bf 1)} Substituting ${1\over\sqrt{v^2+1}}=
  \int_0^{
 \infty}J_0(t)e^{-vt}dt$ in $(III.1.13)$, the BCS-equation with-\par 
\item{}out magnetic field reads 
$$1=const\>\lambda \int_0^{\infty} dt\> J_0(t)
   {1\over t} \left( 1-e^{-{\omega_D
   \over\D}t}\right) \eqno (III.2.7)$$
and is the $B\to 0$ limit of the above equation.
\item{\bf 2)} For zero magnetic field, one has a pairing
   between  $(\bk,\uparrow)$ and 
 $(-\bk,\downarrow)$. With magnetic field, linear 
  momentum $\bk$ is no longer 
 an eigenstate, but angular momentum $l=m-n$ is. 
   Then the  $\triangle\sigma^1$-term 
 in $(III.2.4)$ gives a pairing between 
  $(l,\uparrow)$ and $(-l,\downarrow)$. 
\bigskip
\goodbreak 
\noindent {\bf Proof:}  In the case with magnetic field, the graphs
   contributing to 
 $D(\lambda,\mu,B,\D)$  have to be evaluated in $(n,m)$-space
   at $n=n_B$ and $k_0=0$. 
A two legged graph $\bar G$ is expanded as follows 
$$ \bar G(\xi,\xi')=
     e^{i{B\over2}\x{\x'}^{\bot}}G(\xi-\xi')
  =\sum_{n,m=0}^{\infty}\phi_{nm}(z)\phi_{nm}(z')\int{dk_0\over 2\pi} 
   e^{-ik_0(\tau-\tau')} G(n,k_0)\>, $$
$$\eqalign{G(n,k_0)=\int dxdyd\tau\> l_n{\textstyle 
    \left( {Br^2\over 2} \right)}
     e^{ik_0\tau}G(x,y,\tau)
  =\int{d^2\k\over (2\pi)^2} {2\pi\over B} 2(-1)^nl_n{\textstyle
    \left( {2\k^2\over B}
   \right)} G(\k,k_0)\>. \cr}$$
With $\big< k,-k|V|p,-p\big>=\theta(\omega_D-|e(\k)|)
  \theta(\omega_D-|e(\p)|)$, one obtains to first order
\vskip 0.3cm
$$D(\lambda,\mu,B,\D)=-{\lambda\over2} 
   Tr[\sigma^1(\;first\;order\;graphs\;)|_{n=n_B,k_0=0}
    ]=$$
$$  $$
$${\lambda\over2} Tr[\sigma^1\int d|\k|\>|\k| {1\over B}2(-1)^nl_n
   {\textstyle \left( 
   {2\k^2\over B}\right)} \int{d^3p\over (2\pi)^3}
   \big<k,p|V|p,k\big>\sigma^3
   {\bf D}_{\D}(p)\sigma^3|_{n=n_B,k_0=0}]=$$
$$  $$
$$\lambda\int_0^{\infty}\!ds\>\delta_B(s) 
   \int{d^3p\over (2\pi)^3}\theta(\omega_D-
   |\mu s-\mu|)\theta{\textstyle 
  \left(\omega_D-|{\p^2\over 2}-\mu|\right)}
   \sum_{n=0}^{\infty} 2(-1)^n l_n{\textstyle
   \left( {2\p^2\over B}\right)}
    {\D\over p_0^2+
   \vep_n^2+\D^2}\>. $$
$$ $$
Now, using $\delta_B(s)\approx \delta(s-1)$ and
   performing the $p_0$-integral,
  one gets
$$D(\lambda,\mu,B,\D)=
  const\>\lambda \int_{-\omega_D}^{\omega_D}
  d{\textstyle \left( {\p^2\over2}
   -\mu\right)} 
   \sum_{n=0}^{\infty} 2(-1)^nl_n{\textstyle 
   \left( {2\p^2\over B}\right)} {\D\over 
   \sqrt{\vep_n^2+\D^2}} \eqno (III.2.8)$$
To compute the infinite sum, use the fact that the Laplace transform
   of the zeroth 
 Bessel function is $(s^2+1)^{-{1\over2}}$, that is
$$\int_0^{\infty} dt\>J_0(\D t) e^{-|\vep_n|t}
  ={1\over\sqrt{\vep_n^2+\D^2}} \>.
    \eqno (III.2.9)$$
 The resulting sum can be computed using Theorem II.3. 
 One obtains
$$\eqalignno{ \sum_{n=0}^{\infty}& 2(-1)^nl_n{\textstyle 
    \left( {2\p^2\over B}\right)}
    e^{-|\vep_n|t}=D(\p,-t)-D(\p,t)  \cr
 &=\int_0^{\infty}dv\>\delta_B(v) 
   {1\over \ch^2{Bt\over2}} e^{-{\th{Bt\over2}\over
   {B\over2}} | {\p^2\over2} -\mu v | } \left(  e^{\vep_{n_B}t}
  {\textstyle \theta(
   \mu v-{\p^2\over2})} +e^{-\vep_{n_B}t}{\textstyle 
  \theta( {\p^2\over2}
   -\mu v)}  \right)  \cr
 &\phantom{mm}
  +2\left( {e^{-\vep_{n_B}t}\over 1+e^{-Bt}} -{e^{\vep_{n_B}t}
   \over 1+e^{Bt}}\right) 
   (-1)^{n_B}l_{n_B}{\textstyle 
  \left( {2\p^2\over B}\right)}\>. & (III.2.10)
   \cr}$$
Write $\vep_{n_B}=\alpha B$ with
   $\epsilon<\alpha<1-\epsilon$ (see $(II.16$), 
 and again neglect the smearing in the chemical potential.
   Then $(III.2.8,10)$
 yield
$$D(\lambda,\mu,B,\D)=const\>\lambda\D\int_0^{\infty} dt J_0(\D t) \int_{-\omega_D}^{
   \omega_D}d{\textstyle \left( {\p^2\over2}-\mu\right)}
    \Biggl\{  {1\over \ch^2{Bt\over2}}
   e^{-{\th{Bt\over2}\over{B\over2}} | {\p^2\over2}-\mu |} \times$$
$$\left( e^{\alpha Bt}{\textstyle 
   \theta ( \mu-{\p^2\over2})} +e^{-\alpha Bt} 
  {\textstyle  \theta ( {\p^2\over2}-\mu )} \right) +
   {\sh\left[\left({1\over2}-\alpha\right)
  Bt\right]\over \ch {1\over2}Bt} 2(-1)^{n_B} l_{n_B}{\textstyle 
   \left( {2\p^2\over B}\right)} \Biggr\}$$
$$=const\>\lambda\D\int_0^{\infty} dt\>J_0(t)
   \left\{ \ch\alpha{\textstyle {B\over\D}}t 
   {{B\over\D}
   \over \sh{B\over\D}t} \left( 1-e^{-2{\omega_D\over B}
    \th{Bt\over 2\D}}\right) 
  +{c_B\over 2}{B\over \D} {\sh\left[\left({1\over2}-\alpha\right){
     B\over\D}t\right]\over
   \ch {1\over2}{B\over \D}t}\right\}$$
and again $c_B=\int_{1-{\omega_D\over\mu}}^{1+{\omega_D\over\mu}} 
    ds\>\delta_B(s)$ 
may be approximated by one, which gives the 
  stated equation $\blacksquare$
\bigskip
 In order to have the $B=0$ BCS-equation  $(III.2.7)$ a 
 solution $\D$, the exponent ${\omega_D\over \D}$ must be choosen large.
   In  $(III.2.6)$ 
 however, the magnitude of the exponent is determined by the
   ratio  ${\omega_{D}\over
  B}$ because the $\D$ appears in the hyperbolic tangens which is 
 always bounded by 
 one. Thus in order to get a solution $\D$, $B$ has to be
   sufficiently small.  This 
 becomes clear in considering the following curves.
\bigskip
$$(see\;figure\;in\;CMP\;paper)$$
\bigskip
 For the computation of the critical field, let $b={B\over\omega_D},\;h={\omega_D\over
  \D}$ and let $\D\to 0$ or $h\to\infty$. $(III.2.6)$ becomes 
$$1=const\>\lambda \int_0^{\infty}
   d\tau J_0\left( {\tau\over bh}\right) \left\{
   \ch\alpha\tau{1-e^{-{2\over b}\th{\tau\over2}}\over \sh\tau}
 +{1\over2} {\sh\left[\left(
   {1\over2}-\alpha\right)\tau\right]\over \ch {1\over2}\tau} \right\}$$
$$\buildrel h\to \infty\over \to const\>\lambda
   \int_0^{\infty} d\tau \left\{
   \ch\alpha\tau{1-e^{-{2\over b}\th{\tau\over2}}\over \sh\tau}
 +{1\over2}  {\sh\left[\left(
   {1\over2}-\alpha\right)\tau\right]\over \ch {1\over2}\tau} \right\}$$
 and one computes  
$$B_c=const\>\omega_D e^{-{1\over const\>\lambda}}=const\>\D \>.
    \eqno (III.2.12  )$$
\bigskip
\bigskip
\noindent{\klein III.3 The Flow of the Four Point Function}
\bigskip
 In the preceeding paragraph, it has been shown
   that the appearance of the 
  hyperbolic tangens ${2\over B}\th{B\tau\over 2}$ instead of $\tau$ 
  in the magnetic field free propagator is responsible for the 
 existence of a  critical 
  field.
 Thus, one would expect that the flow of the four legged part of the 
  effective potential behaves differently
   than the $B=0$ flow $(III.1.3,4)$
   if one takes 
 the approximation
$$D(\k,\tau)\approx {1\over\ch^2 {B\tau\over2}} e^{-e(\k) 
   {\th  {B\tau\over2}\over
  { B\over2}}}
   [e(\k),\tau] \equiv \tilde D(
  \k,\tau)  \eqno (III.3.1)  $$
for the exact propagator $(II.8,11)$ 
  and neglects the phase factor in $(II.8a)$.
 This is indeed the case. 
\bigskip
\noindent{\bf Lemma III.3.1:} Substituting the $B=0$ propagator $C$ by
   $\tilde D$, the 
 flow equation in the ladder approximation $(III.1.3)$ becomes
$$\lambda_\ell^{(h-1)}(B)=\lambda_\ell^{(h)}(B)+\beta_B^{(h)} \left(
            \lambda_\ell^{(h)}(B)
   \right)^2  \eqno (III.3.2) $$
where contrary to $(III.1.4b)$ the $\beta_B^{(h)}$'s satisfy
$$\sum_{h=-\infty}^0 \beta_B^{(h)}=const\left( {\textstyle \log{1\over B}}
   +const\right)\>. \eqno (III.3.3)$$
\bigskip
\noindent{\bf Proof:} The $\beta^{(h)}$'s become 
$$\beta_B^{(h)}=\int{dp_0d|\p|\over (2\pi)^3} {|\p|\over k_F} \left( 
   |\tilde D^{(\le h)}(p)|^2
  -|\tilde D^{(< h)}(p)|^2 \right)  \eqno (III.3.4)$$
where 
$$\tilde D^{(\le h)}(p_0,\p)=\!\int\! 
   d\tau\, e^{ip_0\tau} {1\over\ch^2 {B\tau
   \over2}} 
   e^{-e(\p) {\th {B\tau\over2}\over {B\over2}}}
   [e(\p),\tau] \rho^2\!\left( 
   M^{-2(h+1)}{e(\p)}^2\right) $$
 and $\rho$ as in $(III.1.4)$. One computes 
$$\int{dp_0\over2\pi}  \left( |\tilde D^{(\le h)}(p)|^2
   -|\tilde D^{(< h)}(p)|^2  \right) 
   \phantom{mmmmmmmmmmmmmmmmmmmmm}$$
$$=\int_0^{2\over B} dv  (1-{\textstyle {B^2\over4}}v^2) e^{-2|e(\p)| v} 
   \left\{ \rho^2\left( M^{-2(h+1)}{e(\p)}^2\right)-\rho^2\left(
      M^{-2h}{e(\p)}^2\right)
   \right\} \>.$$
Hence $(III.3.4)$ gives
$$\beta_B^{(h)}=
  const \int_0^{\infty} dy \int_0^{2\over B}  dv  
   (1-{\textstyle  {B^2\over4}}v^2) 
   e^{-2 |y-\mu| v} 
   \left\{ \rho^2\left( M^{-2(h+1)}(y-\mu)^2\right)-\rho^2\left(
  M^{-2h}(y-\mu)^2\right)
   \right\} \eqno (III.3.5) $$
These $\beta_B^{(h)}$'s show a different
   behaviour than $(III.1.4a,b)$ since 
$$\eqalignno{ \sum_{h=-\infty}^0 \beta_B^{(h)}&=const 
   \int_0^{\infty} dy \int_0^{2\over B}  dv 
   (1-{\textstyle {B^2\over4}}v^2)
    e^{-2 |y-\mu| v} \rho^2(M^{-2}(y-\mu)^2)  \cr
  &=const \int_0^1 du (1-u^2)
   {1-e^{-4{M^2\over B}u}\over u}= const\left( 
  {\textstyle \log{1\over B} }
   +const \right)  \cr}$$
in contrast to $\lim_{h\to -\infty}
    \beta^{(h)}=\beta\;\blacksquare$
\bigskip
 It has been shown in [FT2], 
  that if 
$$\sup_{\ell\ge 0}\left\{ |\lambda_\ell^{(0)}|\right\}   
   \sum_{h=-\infty}^0\beta_B^{(h)}
   \le\gamma<1 \eqno  (III.3.6)$$
then all sequences of $\lambda_\ell^{(h)}(B)$'s generated by the 
  flow equation $(III.3.2)$
 converge irrespective of the sign of $\lambda_\ell^{(0)}$. 
   That is, if  $(III.3.6)$ is 
satisfied, then the normal ground state is stable, no matter
   whether the potential is 
 attractive or repulsive provided $\lambda$ is small$ $  enough.
   Since $\lambda_\ell^{(0)}$ 
 is proportional to $\lambda$ and because of $(III.3.3)$,
   condition $(III.3.6)$ 
 implies 
$$|\lambda(const\>\log{\textstyle 
      {1\over B}}+const)|<1\;\;\;\hbox{or}\;\;\;
      B>const\> e^{-{const\over|\lambda|}}$$
in agreement with $(III.2.12)$.   
\bigskip
\bigskip
\bigskip\bigskip
\bigskip
\bigskip
\noindent{\mittel IV. Perturbation Theory}
\bigskip
\bigskip
In this Section, we  summarize  without proof (for details, see [Le])
 the results concerning perturbation 
 theory of the model  $(I.1,2,3)$
 where
 $V$ is assumed to be 
  a rotation invariant  potential in $L^1(\Bbb R^3)$. 
 Spin indices are neglected. Since one is  interested in bounds
     which are uniform for
  small $B$, the strategy is the 
 same as in the zero magnetic field case. For $B=0$, it is proven
   in [FT1] that 
\item{---} the ultraviolet part is irrelevant,
         that is each graph is bounded by 
         $const^n$ in the ultraviolet regime;
\item{} For the infrared part, one obtains
\item{---} two legged graphs are in general infinite, they
           have to be  renormalized;
\item{---} four legged graphs produce $n!$'s.
\bigskip
\noindent$\phantom{m}$ In the case with magnetic field, one obtains 
         the same  results uniform in $0\le B\le B_0$, that is
\item{---} the ultraviolet part is irrelevant, each graph is bounded by 
         $const^n$ with a $B$-independent constant;
\item{} For the infrared part, one obtains
\item{---} two legged graphs are finite, but they blow up for small $B$,
         so they have
           to be renormalized;
\item{---} the values of all graphs without two legged subgraphs
         converge to the corresponding values of the 
         $B=0$ graphs as distributions. The same
         holds for renormalized graphs 
         if there are two legged subgraphs. Graphs 
         containing four legged subgraphs may be 
         bounded by $const_B^n$, but in the limit this jumps up 
         to $const^n n!$ which is the uniform bound. 
\bigskip
\noindent$\phantom{m}$The ultraviolet and infrared part of the model 
  are defined by the 
  decomposition 
$$ S(\xi,\xi')=e^{i{B\over2}\x{\x'}^{\bot}}\Bigl(D^{(0)}(\xi-\xi')
   +\sum_{j=j_B}^{-1} D^{(j)}(\xi-\xi') \Bigr)$$
where the ultraviolet part is given by 
$$D^{(0)}(\xi)={B\over2\pi} \sum_{n=0}^{\infty}
   \Ln e^{-\vep_n\tau}[\vep_n,\tau]
                h(\vep_n^2) \eqno (IV.1)$$
and the infrared part at scale $j$ is 
$$D^{(j)}(\xi)={B\over2\pi} \sum_{n=0}^{\infty}
   \Ln e^{-\vep_n\tau}[\vep_n,\tau]
                f(M^{-2j}\vep_n^2) \eqno (IV.2)$$
where $h$ is a smooth monotone function obeying
$$h(x)=\cases{ 0 &if $x\le 1$ \cr
               1 &if $x\ge M^2,$ \cr}$$
$M$ is a real number bigger than one and 
$$f(x)=h(x)\left( 1-h(M^{-2}x)\right)=\cases{ h(x) &if $x\le M^2$\cr
                                 1-h(M^{-2}x) &if $x\ge M^2$ \cr}$$
has support in $[1,M^4]$, thus $f(M^{-2j}x)$ forces 
    $M^{2j}\le x\le M^4M^{2j}$ and  
$$1=h(x)+\sum_{j=-\infty}^{-1} f(M^{-2j}x),\;\;\;x>0\>. \eqno (IV.3)$$
$j_B$ is determined by $M^{j_B+2}=\epsilon B,\;\epsilon B\le
   \vep_{n_B}\le
  (1-\epsilon) B$. The basic estimates are given in the following 
\bigskip
\noindent{\bf Lemma IV.1 (Covariance Estimates):} 
\item{\bf a)} There is the decomposition 
 $D^{(0)}(\xi)=D_{reg}^{(0)}(\xi)+D_{sing}^{(0)}(\xi)$  
 where 
$$\eqalignno{ |D_{reg}^{(0)}(\xi)|&\le const
   {1\over 1+r^4} {1\over 1+\tau^2} &
    (IV.4) \cr
 & \cr
 D_{sing}^{(0)}(\xi) &=-\rho(\xi){1\over 2\pi}
   {{B\over2}\over \sh{B\tau\over2}}
  e^{\mu\tau} e^{-{{B\over2}\over\th{B\tau\over2}}
   {r^2\over2}} \theta(\tau), 
   &(IV.5)  \cr} $$
 $\rho\in C_0^{\infty}$ being one for $|\xi|<1$ and zero for $|\xi|>2$. 
\item{\bf b)} Let $j_B\le j\le-1$ and $N,N'\in \Bbb N$ arbitrary.
   Then there are 
 $\mu$ dependent constants $c_2>c_1>0$ and  
 a constant $const=const_{N,N',M,\mu}$ such that 
$$D^{(j)}(\xi)\le const\>\max_{{c_1\over B}\le n\le {c_2\over B}}
  \Bigl\{ \Bigl|\Ln\Bigr|\Bigr\} M^j \R\T \>. \eqno (IV.6)$$
\item{\bf c)} There are the pointwise limits 
$$\lim_{B\to 0} D^{(0)}(\xi)=C^{(0)}(\xi)\>,\;\;\;
  \lim_{B\to 0} D^{(j)}(\xi)=C^{(j)}(\xi)\eqno(IV.7)$$
where 
$$C^{(0)}(\xi)=\int{d^2\k\over(2\pi)^2}\> e^{i(\k\x-k_0\tau)}
   e^{-e(\k)\tau}[e(\k),\tau] h(e(\k)^2) \eqno (IV.8)$$
is the ultraviolet part and
$$C^{(j)}(\xi)=\int{d^2\k\over(2\pi)^2}\> e^{i(\k\x-k_0\tau)}
   e^{-e(\k)\tau}[e(\k),\tau] f(M^{-2j}e(\k)^2) \eqno (IV.9)$$
is the scale $j$ infrared part of the $B=0$ propagator. 
\bigskip
\bigskip
\noindent$\phantom{m}(IV.4)$ is the same
   as the $B=0$ bound and $(IV.5)$ 
  is smaller than the  
 $B=0$ bound which is $-\rho(\xi){1\over 2\pi\tau} e^{\mu\tau}
   e^{-{r^2\over 2\tau}}
   \theta(\tau)$. Thus the fact that the ultraviolet part of
  $(I.1)$ is irrelevant 
 is an immediate consequence of the corresponding
   result of the $B=0$ model. 
 $(IV.6)$ differs from the $B=0$ bound only
   in the factor  
  $\max_{{c_1\over B}\le n\le {c_2\over B}}
   \Bigl\{ \Bigl|\Ln\Bigr|\Bigr\}$ which would be substituted
    in the latter case by 
 $(1+r)^{-{1\over2}}$. However, 
$$ \M\not\le const\>{1\over (1+r)^{1\over2}} \>,\eqno (IV.10)$$
since the estimate fails near the turning point of the 
  Laguerre function where  the
 decay is only $r^{-{1\over3}}$,  
 so the decomposition 
$${1\over (1+r)^{1\over2}}\le \sum_{k=-\infty}^{-1} const\> M^{{1\over2}k}
    e^{-M^k(1+r)} \>, \eqno (IV.11)$$
which is done to estimate the $B=0$ graphs, has to be substituted by a 
 suitable decomposition of the Laguerre function. This
    can be done (see [Le], 
 lemma IV.1.3). The net effect is, that, as in
    the $B=0$ model, the bound on a 
 labelled graph, that is a graph with scales on all lines,  
 in $2+1$ dimensions can be reduced to the one dimensional
   case  where the 
 covariance $C^{(j_\ell)}_\ell$ at scale $j_\ell$,
    $\ell$ being some line of
   the graph, obeys 
$$|C_\ell^{(j_\ell)}(y)|\le M^{{1\over2}j_\ell} g(M^{j_\ell}y)\>,\;\;\;\;
   g\in L^1(\Bbb R)
  \cap L^\infty(\Bbb R)\>.\eqno (IV.12)$$
   The power counting of such graphs is given 
  by the following 
\bigskip
\goodbreak
\noindent{\bf Lemma IV.2 (Power Counting):} Let $G_{2q}$ be 
   a connected   amputated graph
 with $2q$ external legs build up from 
 generalized vertices or subgraphs $I_{2q_v}$ obeying 
$$\N I_{2q_v}\N_\emptyset\equiv 
   \sup_i \sup_{x_i}\biggl\{ \Bigl(\prod_{j\ne i}
  \int  d^dx_j \Bigr) 
   \>|I_{2q_v}(x_1,\cdots,x_{2q_v})| \biggr\} <\infty\>.$$
For $S\subset\{1,\cdots,2q\}\ne\emptyset$ and testfunctions 
    $f_k\in L^1(\Bbb R^d)\cap 
    L^\infty(\Bbb R^d)$,  introduce the norm 
$$\N G_{2q}\N_S=\int \prod_{i=1}^{2q} d^dx_i \prod_{k\in S} |f_k(x_k)|\>
  |G_{2q}(x_1,\cdots,x_{2q})|\>.$$
Suppose each line of the graph has a covariance $C^{(j)}$ with 
$$|C^{(j)}(x)|\le M^{{d\over2}j}\>g(M^jx)\>,\;\;\;g\in L^1(\Bbb R^d)\cap 
  L^\infty(\Bbb R^d)\>.$$
Then there are the following bounds
$$\N G_{2q}\N_\emptyset\le c^{\sum_vq_v-q} 
   \prod_{v\in V} \Bigl( \N I_{2q_v}
   \N_\emptyset 
   M^{{d\over4}(2q_v-4)j} \Bigr) \>M^{-{d\over4}(2q-4)j} \eqno(IV.13)$$
$$\N G_{2q}\N_S\le c^{\sum_vq_v-q} \prod_{v\in V_{int}} 
   \Bigl( \N I_{2q_v}\N_\emptyset
   M^{{d\over4}(2q_v-4)j} \Bigr) \prod_{v\in V_{ext}} 
   \Bigl( \N I_{2q_v}\N_{S_v} M^{{d\over4}(2q_v-|S_v|)j} \Bigr)\>
   M^{-{d\over4}(2q-|S|)j} \eqno(IV.14)$$
where $c=\max\{\|g\|_{L^1},\|g\|_{L^\infty}\}$. Thereby a vertex
   is called external, 
 if at least one of its legs is integrated against a testfunction. 
\bigskip
\bigskip
Iterating $(IV.13,14)$ for different scales,
   one gets a summable decay for 
 $q_v\ge 3$, a marginal situation which
    produce $n!$'s for $q_v=2$ and an 
 exploding factor for $q_v=1$, that is,
   in the case of two legged subgraphs. 
  They have to be renormalized. Since the magnetic field
   propagator  $(ik_0-\vep_n)^{-1}$ 
 has its maximum at $k_0=0$ and $n=n_B$, the local part of a 
  two legged diagram
 $\bar G(\xi_1,\xi_2)=
   e^{i{B\over 2}\x_1 \x_2^{\bot}}G(\xi_1-\xi_2)$  is 
 given by 
$${\bf L}\int d\xi_1 d\xi_2 \bar G(\xi_1,\xi_2) 
   \bar\psi(\xi_1)\psi(\xi_2) \;=\>
   G(n=n_B, k_0=0) \int d\xi\> \bar\psi(\xi)\psi(\xi)  \eqno (IV.15)$$
where 
$$G(n,k_0)=\int d\tau \>e^{ik_0\tau}\int d^2r\> \Ln G(r,\tau)\>.
       \eqno (IV.16)$$
Then a renormalized graph $(1-{\bf L})\int d\xi_1 d\xi_2
       \bar G(\xi_1,\xi_2) \bar\psi(\xi_1)\psi(\xi_2)$ has indeed 
  an improved power counting since ([Le], lemma IV.3.6,7) 
$$|G(n,k_0)-G(n_B,0)| \le \left( |k_0|+|\vep_n-
  \vep_{n_B}|\right) \|\,|\xi|G(\xi)\|_{L^1}\le
    M^jM^{-i_G}M^{{4\over3}i_G}
    \eqno (IV.17)$$
which is an improvement of $M^{j-i_G}$ since $j<i_G,$ $i_G$ being 
 the lowest scale of $G$, 
 because the renormalized tree expansion produces
   renormalized subgraphs $RG$  only with 
 scale $i_{RG}>j$ whereas counterterm subgraphs $LG$ have 
  scale $i_{LG}\le j$. To 
 review the formalism of renormalization, see for
  example [FT1] or [FKLT1,2], where 
 an inductive treatment is given. 
\par  
Using $(IV.17)$, one can proof ([Le], lemma IV.3.8) as 
  in the $B=0$ case ([FT2], lemma
 II.2), that a string of two legged subgraphs (renormalized 
  and counterterm) may  
 be substituted by a single covariance.
    Then one can apply the power counting 
 lemma without having $q_v=1$ to obtain
\bigskip
\noindent{\bf Theorem IV.3:} Let $G=G(B)$ be a
  (necessarily connected and amputated) $n$'th order graph 
 with $2q$ external legs contributing to the
    renormalized effective potential 
$${\cal G}(\psi^e,\bar\psi^e)=
   \log{1\over { Z}} \int e^{-\lambda{\cal V}(\psi+
   \psi^e,\bar\psi+\bar\psi^e)+\delta\mu(\lambda,B)\int d\xi\>(\psi+
   \psi^e)(\xi)(\bar\psi+\bar\psi^e)(\xi) } d\mu_S(\psi,\bar\psi)\>.$$
Then there is a constant independent of $B$ such that 
$$\N G(B)\N_{\{1,\cdots,2q\}}\le n!\,const^n
    |\lambda|^n \prod_{v\in V_{int}}
   \N V\N_\emptyset\prod_{v\in V_{ext}}\N V\N_{S_v}\>.$$
Furthermore, 
$$\lim_{B\to 0}\N G(B)\N_{\{1,\cdots,2q\}}
    =\N G(0)\N_{\{1,\cdots,2q\}}\>.$$
\vfill
\eject
\noindent {\mittel References}
\bigskip
\bigskip
\bigskip
\item{[FT1]} J. Feldman and E. Trubowitz, Perturbation Theory
             for Many Fermion Systems,
             Helvetica Physica Acta 63, 1990, 156-260.
\item{[FT2]} J. Feldman and E. Trubowitz, The Flow of an 
             Electron-Phonon System to 
             the Superconducting State, Helvetica Physica Acta 64, 
             1991, 214-357. 
\item{[FKLT1]} J. Feldman, H Kn\"orrer, D. Lehmann, E. Trubowitz,
             Fermi Liquids in Two Space Dimensions; extended 
             version of the lectures 
             given by Feldman and Lehmann at the workshop 
             ``Constructive Results 
             in Field Theory, Statistical
             Mechanics and Solid State Physics", 
             Ecole Polytechnique, Palaiseau, France, July 25-27, 1994,
             to appear.
\item{[FKLT2]} J. Feldman, H Kn\"orrer, D. Lehmann 
             and E. Trubowitz, in preparation.
\item{[FMRT]}J. Feldman, J. Magnen, V. Rivasseau, E. Trubowitz,
             An Infinite Volume 
             Expansion for Many Fermion Greens Functions, Helvetica
             Physica Acta 65, 
             1992, 679.
\item{[FW]}  A.L. Fetter and J.D. Walecka, Quantum Theory of
             Many-Particle Systems, 
             McGraw-Hill, 1971.
\item{[GR]}  I.S. Gradshteyn, I.M. Ryzhik, Table of Integrals,
             Series and Products, 
             Academic Press, 1965.
\item{[Le]}   D. Lehmann, A Microscopic Derivation of the 
             Critical Magnetic Field in a 
             Superconductor, Thesis, ETH Z\"urich, 1994.

\end